\documentclass[pre,twocolumn]{revtex4-1}

\usepackage[colorlinks=true,urlcolor=blue,citecolor=blue,linkcolor=blue]{hyperref}
\usepackage{tikz}
\usetikzlibrary{trees}

\graphicspath{{../figures/}}

\begin{document}

\tikzstyle{every node}=[draw=black,thick,anchor=west]
\tikzstyle{file}=[draw=green]
\tikzstyle{directory}=[draw=blue,fill=gray!25]
\tikzstyle{optional}=[dashed,fill=gray!50]

\title{A model project for reproducible papers: \\ critical
  temperature for the Ising model on a square lattice}

\author{M. Dolfi}
\affiliation{Theoretische Physik, ETH Zurich, 8093 Zurich, Switzerland}
\author{F. Chirigati}
\affiliation{New York University (NYU), New York, USA}
\author{J. Freire}
\affiliation{New York University (NYU), New York, USA}
\author{J. Gukelberger}
\affiliation{Theoretische Physik, ETH Zurich, 8093 Zurich, Switzerland}
\author{A. Hehn}
\affiliation{Theoretische Physik, ETH Zurich, 8093 Zurich, Switzerland}
\author{J. Imri\v{s}ka}
\affiliation{Theoretische Physik, ETH Zurich, 8093 Zurich, Switzerland}
\author{K. Pakrouski}
\affiliation{Theoretische Physik, ETH Zurich, 8093 Zurich, Switzerland}
\author{T. F. R{\o}nnow}
\affiliation{Theoretische Physik, ETH Zurich, 8093 Zurich, Switzerland}
\author{D. Shasha}
\thanks{Authors are listed in alphabetical order.}
\affiliation{New York University (NYU), New York, USA}
\author{M. Troyer}
\affiliation{Theoretische Physik, ETH Zurich, 8093 Zurich, Switzerland}
\author{I. Zintchenko}
\thanks{Authors are listed in alphabetical order.}
\affiliation{Theoretische Physik, ETH Zurich, 8093 Zurich, Switzerland}

\begin{abstract}
  In this paper we present a simple, yet typical simulation in
  statistical physics, consisting of large scale Monte Carlo
  simulations followed by an involved statistical analysis of the
  results. The purpose is to provide an example publication to explore
  tools for writing reproducible papers. The simulation estimates the
  critical temperature where the Ising model on the square lattice
  becomes magnetic to be $T_c/J=2.26934(6)$ using a finite size
  scaling analysis of the crossing points of Binder cumulants. We
  provide a virtual machine which can be used to reproduce all figures
  and results.
\end{abstract}

\maketitle

\section{Introduction}

The principle that scientific publications have to be reproducible is
a cornerstone of modern science. A theoretical paper typically
contains all the steps required to follow the arguments and arrive at
the final result. Experimental papers usually go to great lengths in
describing the most important details and keep track of workflow in
sacrosanct lab notebooks. In computational science the goal of
reproducibility is harder to achieve. Reasons are the complexity of
computer systems, codes, and analysis procedures and the absence of
well established community guidelines and wide-spread tools for
reproducibility.

To focus the discussion on comparison of procedures and tools, we
picked one simple, yet representative example of a computer simulation
in statistical physics, calculation of the critical temperature of the
classical Ising model. The current manuscript already contains
sufficient details, codes, and scripts to reproduce all the presented
numerical results and figures. However, reaching this level of
reproducibility required efforts that went far beyond simply obtaining
the results. Our goal for future work is to use this paper as an
example for a discussion on how reproducibility can best be achieved.

The total computation time for a small system size is in the order of
several hours on a single core producing several MB of raw output
data. Including large system sizes increases the accuracy of the
results, but also the runtime and the amount of data produced and one
might need to use large computer clusters. This allows to explore the
scalability of tools for reproducibility in computational science.

\section{Critical temperature for the Ising model on a square lattice}
\subsection{Model}

The Ising model dates back to 1920 when it was proposed by Wilhelm
Lenz as a mathematical model for ferromagnetism and first analytically
solved by his student Ernst Ising in one dimension~\cite{IsingModel}.
We will consider the two-dimensional Ising model on a square lattice
of size $L\times L$ with periodic boundary conditions. It is described
by the energy function
\begin{equation}
  H = -J\sum_{\langle i,j \rangle} \sigma_i \sigma_j.
\end{equation}
where the sum is over nearest neighbours, $\sigma_i \in \{ -1, +1 \}$
is the spin on site $i$ and $J$ is the coupling strength. In the
following we will focus on the ferromagnetic model with $J>0$.

At high temperatures $T$ the system is unordered.
Each spin $\sigma_i$ has a random orientation and no magnetisation
\begin{equation}
 \left< |m| \right> =
 \left< \left|\frac{1}{L^2}\sum_i\nolimits \sigma_i\right| \right> = 0
\end{equation}
can be observed for an infinite system $L\to \infty$.
However, below a critical temperature $T_c$ a magnetic order is
observed --- the spins $\sigma_i$ align in one direction and the model
shows a finite magnetisation $\left<|m|\right> \neq 0$.
For the square lattice without an external field $T_c$ is known
analytically~\cite{IsingModel2DDerivation}
\begin{equation}
  T_c=\frac{2 J} {\ln(1 + \sqrt{2})} \approx 2.269185.
\end{equation}
and provides a benchmark for our results.

\subsection{Methods}

A Monte Carlo simulation using Wolff cluster
updates~\cite{WolffAlgorithm} is used to construct new system
configurations, employing the MT19937 Mersenne Twister pseudo random
number generator \cite{RNG}. For each parameter set more than
$1280000$ measurements are performed after discarding $10\%$ additional
Wolff updates for thermalization; error estimates are done with
binning analysis \cite{dogs-fleas}.

The critical temperature can be
roughly estimated from the connected susceptibility
\begin{equation}
  \label{eq:suscept}
  \left<\chi\right>_\beta = \beta L^2 \left(\left<m^2\right> - \left<|m|\right>^2\right),
\end{equation}
where $\beta=\frac{1}{k_BT}$ and $k_B$ is the Boltzmann constant. The
average is taken over different
configurations. $\left<\chi\right>_\beta$ has a peak around $T_c$
\cite{BinderCumulants}, which gives a first rough estimate.

The Binder cumulant
\begin{equation}
  U_2 = \langle m^2 \rangle / \langle |m| \rangle^2
  \label{eq:binder}
\end{equation}
provides a more accurate technique to extract the critical
temperature. For different system sizes the temperature dependence of
$U_2$ is expected to cross at different points. The crossing points
can be shown to follow
\begin{equation}
  T_c^*(L) = T_c^* + AL^{-1/\nu}
\end{equation}
where the critical exponent $\nu = 1$ in two dimensions
\cite{CriticalExponents}. To extract $T_c^*$ for an infinite system we
now fit the positions of the crossing points between system sizes $L$
and $L/2$, respectively, using a least-squares fit weighted with the
size of the error bars at each system size, minimizing $\chi^2=\sum_L
(T_L - T^l_L)^2 / \zeta_L^2$, where $T_L$ is the crossing point
temperature for the systems with sizes $L$ and $L/2$, $\zeta_L$ is its
standard error, and $T^l_L$ is the value of the linear fitting
function. A Jacknife analysis with at least $78$ bins is used to
estimate the errors of the Binder cumulants and their crossings.

\subsection{Results and Discussion}

\begin{figure}
  \centering
  \includegraphics[width=0.5\textwidth]{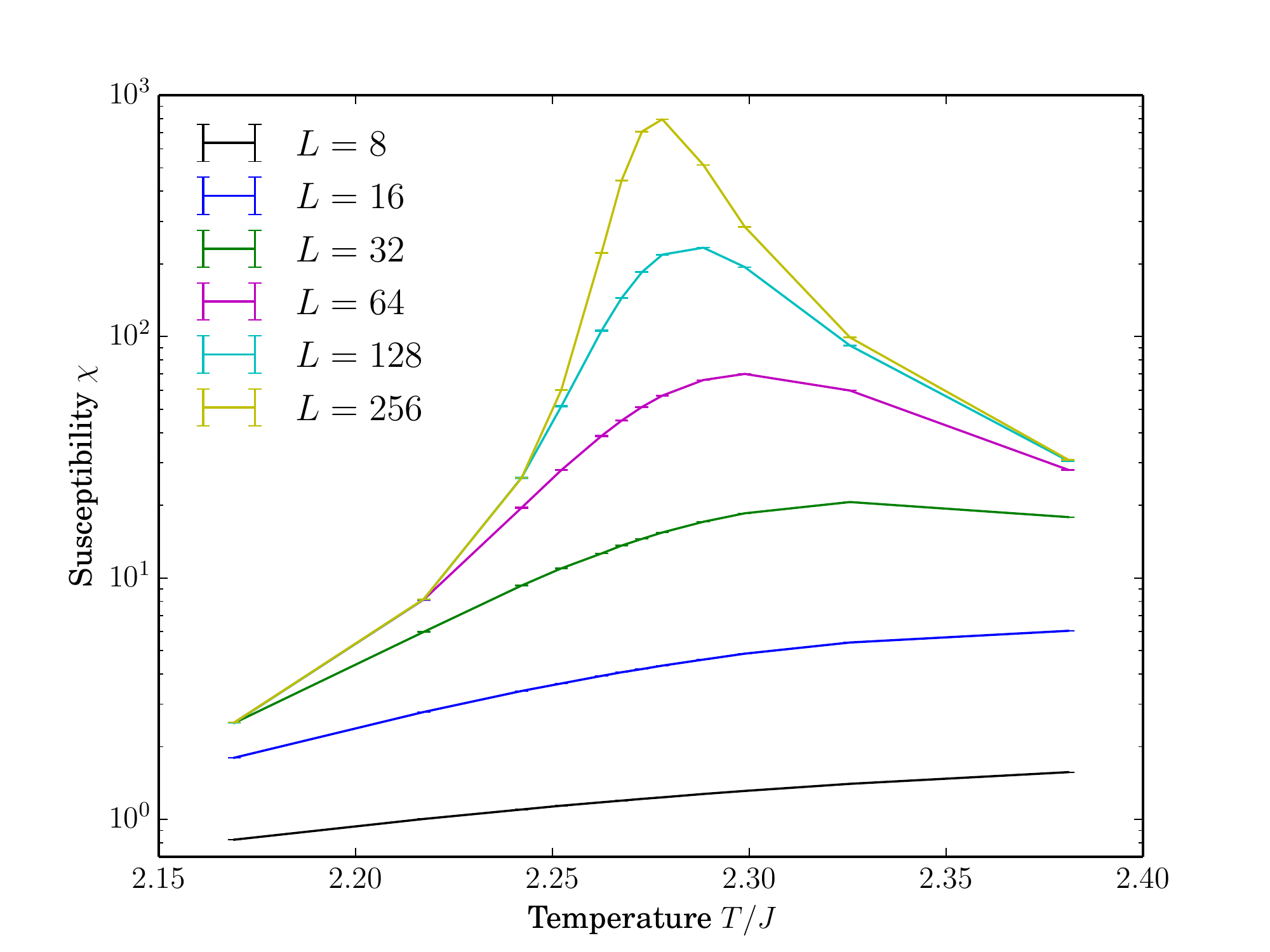}
  \caption{Temperature dependence of the susceptibility for different
    system sizes $L$.}
  \label{fig:susceptibility}
\end{figure}

\begin{figure}
  \begin{center}
    \includegraphics[width=0.5\textwidth]{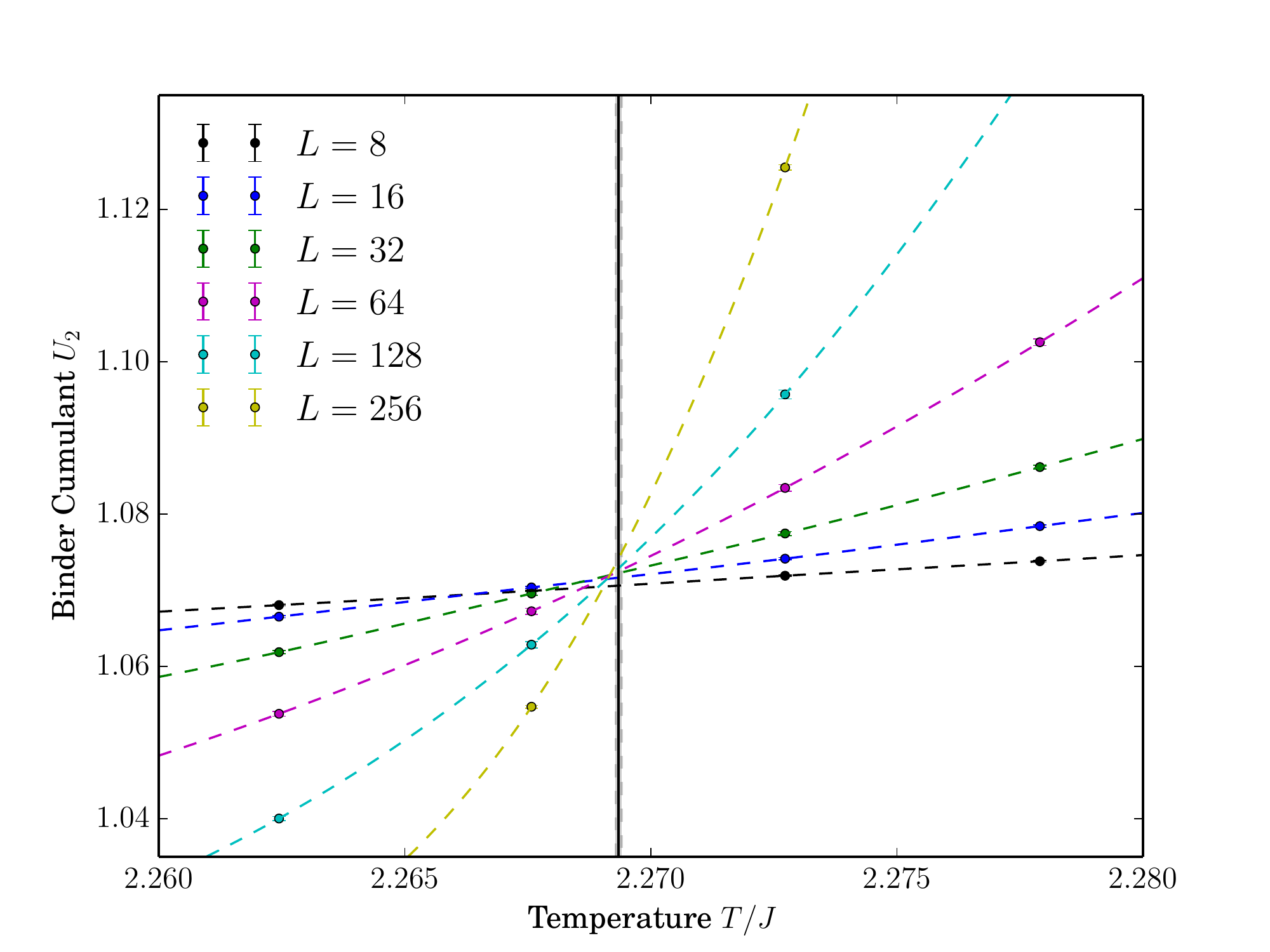}
    \caption{Temperature dependence of the Binder cumulants for
      different system sizes. Vertical line and the grey area around
      it indicate our estimate for critical temperature and for the
      error respectively. Dashed lines are  fits to a cubic polynomial.}
    \label{fig:bindercross}
  \end{center}
\end{figure}

\begin{figure}
  \begin{center}
    \includegraphics[width=0.5\textwidth]{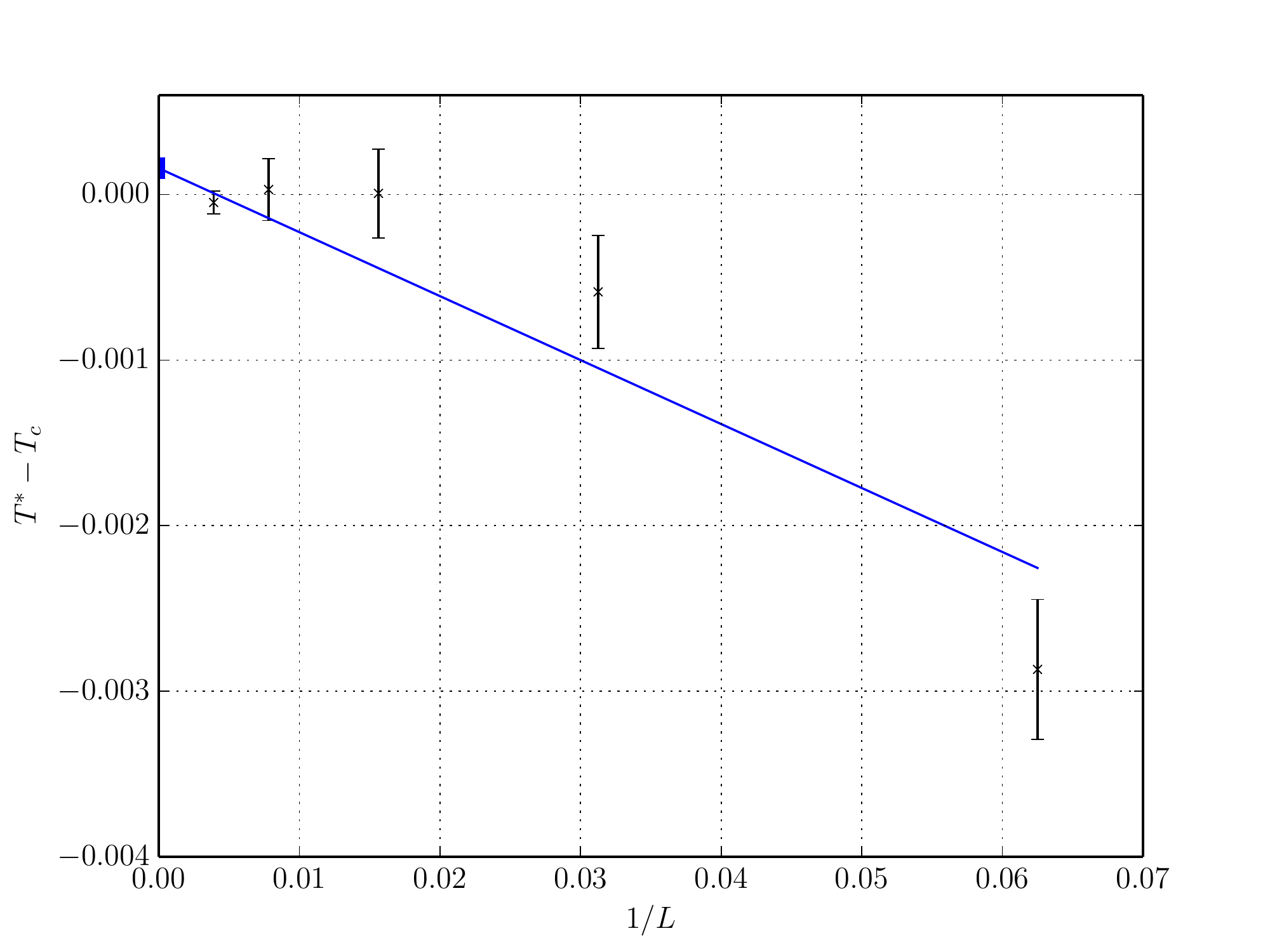}
    \caption{Finite size scaling of the Binder cumulant crossing
      points. The dependence can be shown to satisfy $T_c^*(L) = T_c^* +
      AL^{-1/\nu}$ and $T_c^*/J=2.26934(6)$ is extracted from a weighted
      least-squares fit as described in the text.}
    \label{fig:sizescal}
  \end{center}
\end{figure}

\begin{figure}
  \begin{center}
    \includegraphics[width=0.5\textwidth]{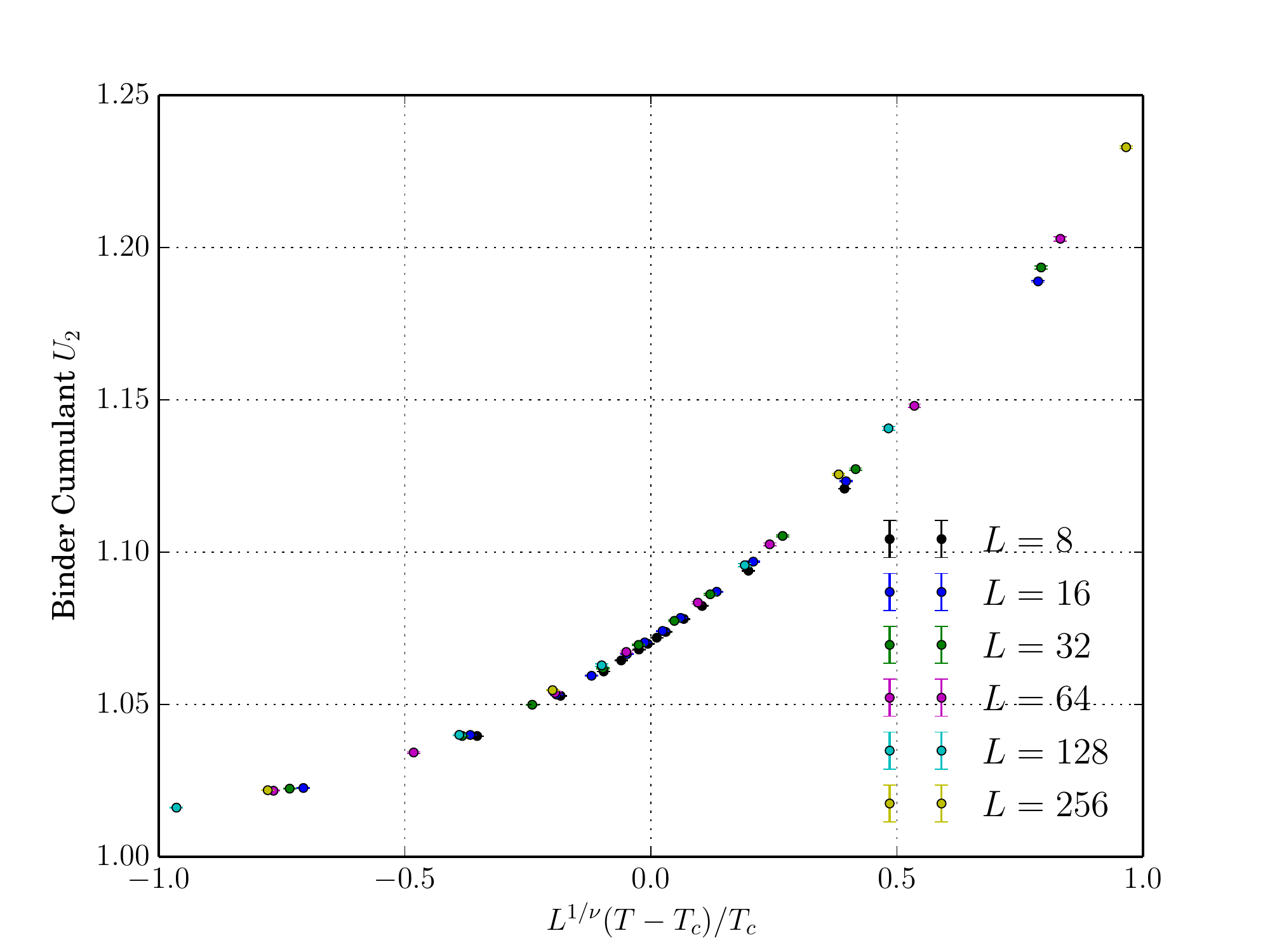}
    \caption{Data collapse of Binder cumulant. $T_c^*$ is obtained from
      Fig.~\ref{fig:sizescal}.}
    \label{fig:bindercollapse}
  \end{center}
\end{figure}

Figure \ref{fig:susceptibility} shows the connected susceptibility
defined in Eq.~\ref{eq:suscept} as a function of temperature for
system sizes $L=8,16,\ldots, 256$. It peaks around $T \approx 2.275J$,
which gives a first estimate for the critical temperature.

Figure \ref{fig:bindercross} shows the intersections of Binder
cumulants $U_2$ (\ref{eq:binder}) with cubic interpolation between all
temperatures within each system size. For each pair of consecutive
system sizes we identify the temperature at the crossing point. This
temperature is plotted in Fig.~\ref{fig:sizescal} as a function of the
larger system size in each pair. By extrapolating to the limit $1 / L
\rightarrow 0$, a more accurate estimate for the critical temperature
$T_c^* = 2.26934(6) J$ is obtained. To check the validity of this
value, we plot a data collapse of the Binder cumulant $U_2$ for
different system sizes versus $L^{1/\nu}(T-T_c)/T_c$
(Fig.~\ref{fig:bindercollapse}), which according to finite size
scaling should satisfy
\begin{equation}
  U_2 = \mathcal{F}(L^{1/\nu}(T-T_c)/T_c),
  \label{eq:binderScaling}
\end{equation}
where $\mathcal{F}$ is a universal function. The curves are indeed on
top of one another in the vicinity of our estimate for the critical
point $T_c^*$. The reader is invited to use our scripts to test
different values of $T_c^*$ and $\nu$ in the data collapse
(appendix \ref{sec:data_collapse}).

\subsection{Conclusion}

Our final estimate for the critical temperature $T_c^*/J = 2.26934(6)$
is consistent with the analytically known value, which is a good check
for the correctness of the analysis in this demonstration
paper. Instructions to reproduce all figures and results are provided
in appendix \ref{sec:setup}. As one might expect, the effort involved
in making this paper reproducible was much larger than the effort for
creating the first results. An important lesson learned from this
project is that it will be important to develop best practices and
better tools to more easily make simulations reproducible. This
prototypical simulation can be a good non-trivial, yet not too complex
example of a test project to explore the capabilities of various
tools.

\appendix

\section{Reproducing our setup}
\label{sec:setup}

This appendix contains detailed instructions for reproducing the
numerical results presented in this paper. Additionally we point out
parameters the reader may want to change for checking the robustness
of our analysis. We provide a virtual machine
image~\footnote{\url{{http://archive.comp-phys.org/provenance_challenge/provenance_machine.ova}}},
which can be imported into the open source software
VirtualBox~\footnote{\url{http://www.virtualbox.org}} and used to
recreate all our results from scratch or run additional
simulations. The machine can be accessed by logging in as user ``Mad
Scientist'' (password: \verb|ilovemath|), opening a terminal and
changing into the directory \verb|~/ising_project|.

\subsection{Running the code}
\label{sec:running}

There are three versions of the simulation that can be used
interchangeably to create the output data:
\begin{itemize}
\item{\verb|ising_single|}: single-threaded simulation
\item{\verb|ising_threaded|}: parallel simulation on shared memory
\item{\verb|ising_mpi|}: parallel simulation on distributed memory
\end{itemize}
The simulation is started by executing the script
\begin{verbatim}
$ python src/runising.py <sim_executable>   \
      <length> <measurements>               \
      <beta> [<beta_2> ... <beta_n>]
\end{verbatim}
where \verb|<sim_executable>| is the executable name, \verb|<length>|
is the linear dimension of the system $L$, \verb|<measurements>| is
the number of Monte-Carlo measurements and \verb|<beta>| is a list of
the inverse temperatures $\beta$. The script will write a log file
along with the output of the simulation, which contains important
provenance information such as the time when the simulation was
started or the hostname of the machine on which the simulation was
run. In addition we provide a convenience script
\verb|python simulate_all.py| to recreate all numerical data presented
in this paper. This will take a few hundred CPU hours. This script
can easily be customised to calculate different system sizes, improve
statistics by increasing the number of sweeps, or extend the
temperature range. By default this script will use the multi-threaded
version of the Ising Monte-Carlo code. The single-threaded or MPI
version can be selected by changing the variable \verb|executable| in
these scripts.

\subsection{Data evaluation}
\label{sec:analysis}

Analyzing the raw data we calculate the estimate for $T_c$ and
generate the figures in both portable document and text formats for
easy access to the numerical values of the data shown. All data
analysis is performed by Python scripts in the \verb|figures|
directory, which operate on raw data stored in the \verb|data|
directory.
The scripts \verb|susceptibility.py|, \verb|binder_cumulant.py| and
\verb|binder_collapse.py| create all figures needed for the paper in
separate subdirectories while \verb|parms.py| contains parameters that
control the behaviour of these scripts.  In detail the reader needs to
execute the following sequence of commands:
\\
\verb|    $ cd figures| \\
\verb|    $ python susceptibility.py| \\
\verb|    $ python binder_cumulant.py| \\
\verb|    $ python binder_collapse.py| \\
For each figure a corresponding directory is created, containing a PDF
file and accompanying text file.

The estimate of the critical temperature $T_c^*$ is printed to the
standard output by \verb|binder_cumulant.py| and has to be manually
copied into \verb|parms.py| whenever the raw data or evaluation has
been changed because it is used as an input for the data collapse.

\subsection{Suggestions for further analysis}
\label{sec:further}

Here we provide some ideas for how one might want to change the
evaluation parameters in \verb|parms.py| for checking our evaluation
procedure.  After editing a parameter, the evaluation needs to be
rerun as described above.

\subsubsection{Determination of Binder crossings}

The fitting function used to determine the crossing points between the
Binder cumulant curves of different system sizes affects our final
$T_c$ estimate. Changing the parameter \verb|binder_crossing_fit_kind|
from \verb|'cubic'| to, e.g., \verb|'linear'| will change the crossing
temperatures plotted in Fig.~\ref{fig:sizescal} and hence the final
estimate $T_c^*$ calculated by \verb|binder_cumulant.py|. The reader
may get an impression of the systematic error connected to this choice
by checking how $T_c^*$ and its error estimate change when the fit
parameter is changed.  A change in the critical temperature estimate
$T_c^*$ will also change the data collapse
Fig.~\ref{fig:bindercollapse} (when the new value output by
\verb|binder_cumulant.py| is copied into \verb|parms.py|).

\subsubsection{Finite size fitting}

Another choice affecting the final result is the range of system sizes
that is used for the extrapolation to the thermodynamic limit as shown
in Fig.~\ref{fig:sizescal}.  The reader may evaluate the effect of
changing the values of the parameters \verb|finite_size_min_L| and
\verb|finite_size_max_L| on $T_c^*$ and its error estimate as printed
by \verb|binder_cumulant.py|.

\subsubsection{Data collapse}
\label{sec:data_collapse}

For a good estimate of the critical temperature and exponent, all data
points in Fig.~\ref{fig:bindercollapse} should lie on a single curve.
By changing the values of \verb|Tc| and \verb|nu| one can explore the
regime of critical temperature $T_c^*$ and exponent $\nu$ where there
is still good data collapse.

\bibliographystyle{apsrev4-1}
\bibliography{paper}

\onecolumngrid

\section{Overview of the source tree}

\begin{tikzpicture}[%
  grow via three points={one child at (0.5,-0.7) and
    two children at (0.5,-0.7) and (0.5,-1.4)},
  edge from parent path={(\tikzparentnode.south) |- (\tikzchildnode.west)}]
  \node {\textbf{Project}}
  child { node [file] {\textbf{simulate\_small.py:} run the simulations for small system sizes (L=8,16,32,64)}}
  child { node [file] {\textbf{simulate\_all.py:} run all simulations reported in the paper (L=8,16,32,64,128,256)}}
  child { node [directory] {\textbf{bin:} simulation executables}}
  child { node [directory] {\textbf{data:} data and log files (recreate with simulate\_all.py}}
  child { node [directory] {\textbf{figures:} plots used in the paper}
    child { node [file] {\textbf{parms.py:} evaluation parameters}}
    child { node [file] {\textbf{susceptibility.py, binder\_cumulant.py, binder\_collapse.py:} scripts creating the figures}}
    child { node [directory] {\textbf{fig\_*:} one directory per figure}
      child { node [file] {\textbf{fig\_*.pdf:} figure in pdf-format}}
      child { node [file] {\textbf{fig\_*.txt:} data shown in the figures in plain text format}}
    }
    child [missing] {}
    child [missing] {}
    child { node [directory] {\textbf{pytools:} auxiliary scripts}
      child { node [file] {\textbf{*.py}}}
      }
  }
  child [missing] {}
  child [missing] {}
  child [missing] {}
  child [missing] {}
  child [missing] {}
  child [missing] {}
  child { node [directory] {\textbf{src:} source files }
    child { node [file] {\textbf{CMakeLists.txt:} cmake project file for building the simulation binary}}
    child { node [file] {\textbf{ising.cpp, ising.hpp:} \texttt{C++} source code of the Monte Carlo simulation}}
    child { node [file] {\textbf{mpi.cpp:} MPI version}}
    child { node [file] {\textbf{threads.cpp:} multi-threaded version}}
    child { node [file] {\textbf{runising.py:} a script to run the simulations.}}
  }
  child [missing] {}
  child [missing] {}
  child [missing] {}
  child [missing] {}
  child [missing] {}
  child { node [directory] {\textbf{paper:} manuscript files }
    child { node [file] {\textbf{paper.tex:} main}}
    child { node [file] {\textbf{paper.bib:} bibliography}}
  }
  ;

\end{tikzpicture}

\end{document}